\documentclass[prd,12pt,letterpaper,nofootinbib]{revtex4}
\usepackage{sidecap}
\usepackage{natbib}
\usepackage{amssymb,graphicx}

\newcommand{\Mpc}{\mbox{ Mpc}}

\newcommand{\Mpcinv}{\mbox{ Mpc$^{-1}$}}

\newcommand{\sqm}{\mbox{ m$^2$}}

\newcommand{\kel}{\mbox{ K}}
\newcommand{\mkel}{\mbox{ mK}}

\newcommand{\hr}{\mbox{ hr}}

\newcommand{\MHz}{\mbox{ MHz}}

\newcommand{\bxhi}{\bar{x}_{\rm HI}}
\newcommand{\xhi}{x_{\rm HI}}

\newcommand{\dtb}{\delta T_b}
\newcommand{\bdtb}{\bar{\delta T}_b}

\newcommand{\lya}{Ly$\alpha$ }

\newcommand{\deriv}{{\rm d}}
\newcommand{\bq}{\begin{equation}}
\newcommand{\eq}{\end{equation}}
\newcommand{\bqa}{\begin{eqnarray}}
\newcommand{\eqa}{\end{eqnarray}}
\def\VEV#1{\left\langle #1\right\rangle} 

\begin{document}

\title{Astrophysics from the Highly-Redshifted 21 cm Line}

\author{Steven~R.~Furlanetto$^1$, 
Adam~Lidz$^2$, 
Abraham~Loeb$^2$, 
Matthew~McQuinn$^2$, 
Jonathan~R. Pritchard$^2$, 
James~Aguirre$^3$,
Marcelo~A.~Alvarez$^4$,
Donald~C.~Backer$^5$, 
Judd~D.~Bowman$^6$, 
Jack~O.~Burns$^7$, 
Chris~L.~Carilli$^8$, 
Renyue~Cen$^9$, 
Asantha~Cooray$^{10}$, 
Nickolay~Y.~Gnedin$^{11}$, 
Lincoln~J.~Greenhill$^2$,
Zoltan~Haiman$^{12}$, 
Jacqueline~N.~Hewitt$^{13}$,
Christopher~M.~Hirata$^6$,
Joseph~Lazio$^{14}$, 
Andrei~Mesinger$^9$, 
Piero~Madau$^{15}$,  
Miguel~F.~Morales$^{16}$,
S.~Peng~Oh$^{17}$, 
Jeffrey~B.~Peterson$^{18}$, 
Ylva~M.~Pihlstr{\" o}m$^{19}$, 
Paul~R.~Shapiro$^{20}$, 
Max~Tegmark$^{13}$,
Hy~Trac$^2$, 
Oliver~Zahn$^5$, 
\& Matias~Zaldarriaga$^2$}

\affiliation{$^1$University of California, Los Angeles; sfurlane@astro.ucla.edu; (310) 206-4127
\\ $^2$Harvard-Smithsonian Center for Astrophysics
\\ $^3$University of Pennsylvania
\\ $^4$Stanford University
\\ $^5$University of California, Berkeley
\\ $^6$California Institute of Technology
\\ $^7$University of Colorado, Boulder
\\ $^8$National Radio Astronomy Observatory
\\ $^9$Princeton University
\\$^{10}$University of California, Irvine
\\$^{11}$Fermi National Accelerator Laboratory and University of Chicago
\\$^{12}$Columbia University
\\$^{13}$Massachusetts Institute of Technology
\\$^{14}$Naval Research Laboratory
\\ $^{15}$University of California, Santa Cruz
\\$^{16}$University of Washington
\\ $^{17}$University of California, Santa Barbara
\\ $^{18}$Carnegie Mellon University
\\ $^{19}$University of New Mexico
\\ $^{20}$University of Texas-Austin
}

\maketitle

\vskip -0.3in

\noindent
Submitted for consideration by  the Astro2010 Decadal Survey Science Frontier Panel \\ \emph{Galaxies across Cosmic Time}

\begin{center}
\includegraphics[width=3.8in]{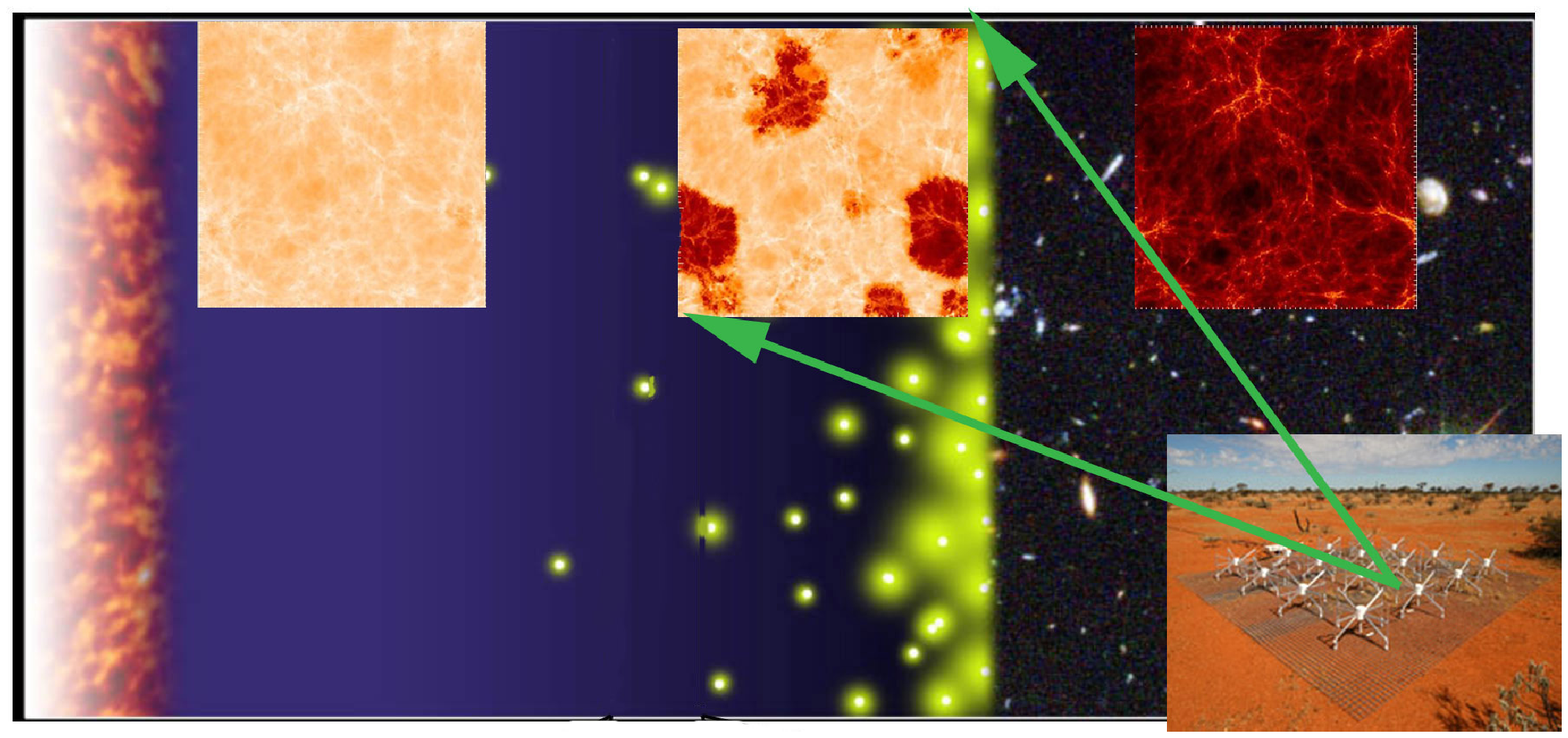}
\end{center} 

\vskip -0.2in

\noindent
The history of the Universe from recombination to the present, with example 21 cm signals and the Murchison Widefield Array overlaid.  Credit: S.~Furlanetto, J.~Lazio, and http://www.mwatelescope.org.

\newpage

\setcounter{page}{1}

\section{Introduction} \label{intro}

Over the past several decades, astrophysicists have pushed the ``high-redshift frontier," where the most distant known galaxies and quasars reside, farther and farther back -- now encompassing star-forming galaxies at $z \sim 7$ \cite{bouwens08} and billion-solar mass black holes shining as quasars at $z \sim 6.5$ \cite{fan06}.  Two strategies can extend these efforts even farther over the next decade. The first, direct observations with large ground- and space-based near-infrared telescopes, will teach us about processes internal to these objects.  But a second method promises a beautiful complement to these direct probes:  low-frequency radio arrays using the 21 cm hyperfine line of neutral hydrogen to explore this era through the galaxies' indirect effects on the intergalactic medium (IGM).  Here we will describe the key astrophysical questions that this unique tool can address:  {\bf What were the properties of high$-z$ galaxies? How did they affect the Universe around them?}

\section{Scientific Context} \label{basic}

\begin{figure}
\includegraphics[width=6in]{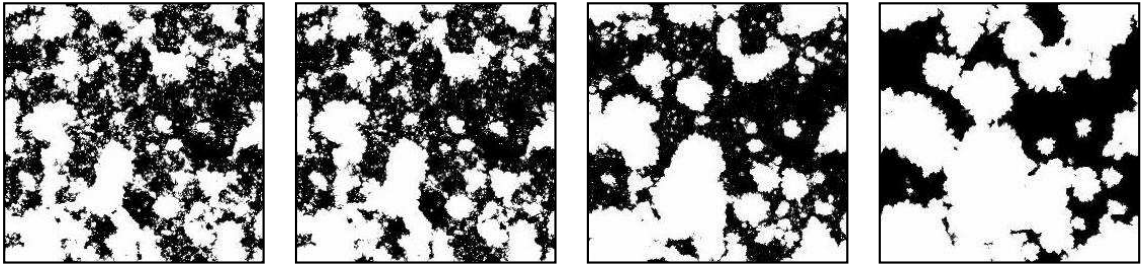}
\caption{Maps ($\approx 0.6^{\circ}$ or $94$ comoving Mpc across) of ionized (white) and neutral (black) gas during reionization, showing the features visible in 21 cm maps.  From left to right, the panels assume that increasingly massive galaxies drive reionization (all have $\xhi \approx 0.5$ and take $z \approx 7.7$).  From \cite{mcquinn07}.}
\label{fig:pics}
\end{figure}

The 21 cm brightness temperature of an IGM gas parcel at a redshift $z$, relative to the cosmic microwave background (CMB), is \cite{madau97, furl06-review}
\bq
\dtb \approx 25\;\xhi(1+\delta) \, \left( { 1+z \over 10} \right)^{1/2}\, \left[1-\frac{T_{\rm CMB}(z)}{T_S}\right] \, \left[ \frac{H(z)/(1+z)}{\deriv v_\parallel/\deriv r_\parallel} \right] \mkel,
\label{eq:dtb}
\eq
where $\xhi$ is the neutral fraction, $\delta = \rho/\VEV{\rho}-1$ is the fractional IGM overdensity in units of the mean, $T_{\rm CMB}$ is the CMB temperature, $T_S$ is the spin (or excitation) temperature of this transition, $H(z)$ is the Hubble constant, and $\deriv v_\parallel/\deriv r_\parallel$ is the line-of-sight velocity gradient.

All four of these factors contain unique astrophysical information.  The dependence on $\delta$ traces the development of the cosmic web \cite{scott90}, while the velocity factor sources line-of-sight ``redshift-space distortions" that separate aspects of the cosmological and astrophysical signals \cite{barkana05-vel}.  The other two factors depend strongly on the ambient radiation fields in the early Universe:  the ionizing background for $\xhi$ and a combination of the ultraviolet background (which mixes the 21 cm level populations through the Wouthuysen-Field effect \cite{wouthuysen52, field58}) and the X-ray background (which heats the gas \cite{chen04}) for $T_S$.  Fig.~\ref{fig:pics} shows how $\xhi$ influences the 21 cm signal:  ionized regions appear as ``holes" in the $\sim 20 \mkel$ background of neutral gas.

\begin{SCfigure}
\centering
\includegraphics[scale=0.35]{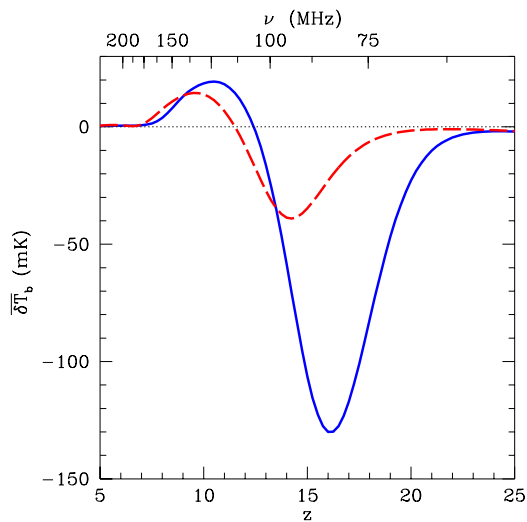}
\caption{Fiducial histories of the sky-averaged 21 cm brightness temperature, $\bdtb$.  The solid blue curve uses a typical Population II star formation history, while the dashed red curve uses only very massive Population III stars.  Both fix reionization to end at $z_r \approx 7$. From \cite{furl06-glob}.}
\label{fig:global}
\end{SCfigure}

Fig.~\ref{fig:global} shows two example scenarios for the sky-averaged 21 cm brightness temperature (see also \cite{gnedin04}).  In the solid curve (a very typical model), we assume that high-$z$ star formation has similar properties to that at lower redshifts.  When the first galaxies form (here at $z \sim 25$), the Wouthuysen-Field effect triggers strong 21 cm absorption \cite{madau97}.  Somewhat later, X-rays -- produced by the first supernovae or black holes -- heat the entire IGM above $T_{\rm CMB}$, turning this absorption into emission -- which fades and eventually vanishes as ionizing photons from the maturing galaxies destroy the intergalactic HI.

The red dashed curve shows a scenario in which very massive Population III stars dominate.  Their reduced efficiency for Wouthuysen-Field coupling (relative to their ionizing emissivity) decreases the strength of the absorption era \cite{furl06-glob}, leading to a markedly different history.  The 21 cm line is clearly sensitive to the basic parameters of high-$z$ star formation.

We have so far illustrated the 21 cm signal through its all-sky background, but of course it actually fluctuates strongly as individual IGM regions grow through gravitational instability and are heated and ionized by luminous sources.  These fluctuations can ideally be imaged, but over the short term statistical measurements are likely to be more powerful  (see \S \ref{mile} below).  Fortunately, as we discuss next (also see Fig.~\ref{fig:pk}), these statistical fluctuations contain an enormous amount of information about this era.

\section{Key Questions} \label{reion}

The rich physics of the 21 cm line allows us to study the over-arching questions of {\bf how the first galaxies evolved and affected the Universe around them}.  In particular, we can approach these questions from the following specific directions.  (Note that related cosmological questions, and more detailed IGM physics, are addressed in the companion white paper ``Cosmology from the Highly-Redshifted 21 cm Transition.")

\underline{\it When did reionization occur?}  One of the most dramatic events in the IGM's history was the reionization of intergalactic hydrogen, most likely via ultraviolet photons from star-forming galaxies.  This event marked the point when the small fraction of matter inside galaxies completely changed the landscape of the diffuse IGM gas (and hence future generations of galaxies), rendering it transparent to high-energy photons and heating it substantially.  The 21 cm background provides the \emph{ideal} probe of reionization.  Its weak oscillator strength (in comparison to Ly$\alpha$) allows us to penetrate even extremely high redshifts.  We can also image it across the entire sky -- instead of only rare, isolated \lya forest lines of sight.  Moreover, unlike the CMB, it is a spectral line measurement, and we can distinguish different redshift slices and study the full history of the ``dark ages" -- extremely difficult even with a ``perfect" CMB measurement \cite{zald08-cmbpol}.  Finally, it directly samples the $ 95\%$ (or more) of the baryons that reside in the IGM, requiring no difficult inferences about this material from the properties of the rare luminous galaxies. 

\begin{figure}[htbp]
\begin{center}
\includegraphics[scale=0.35]{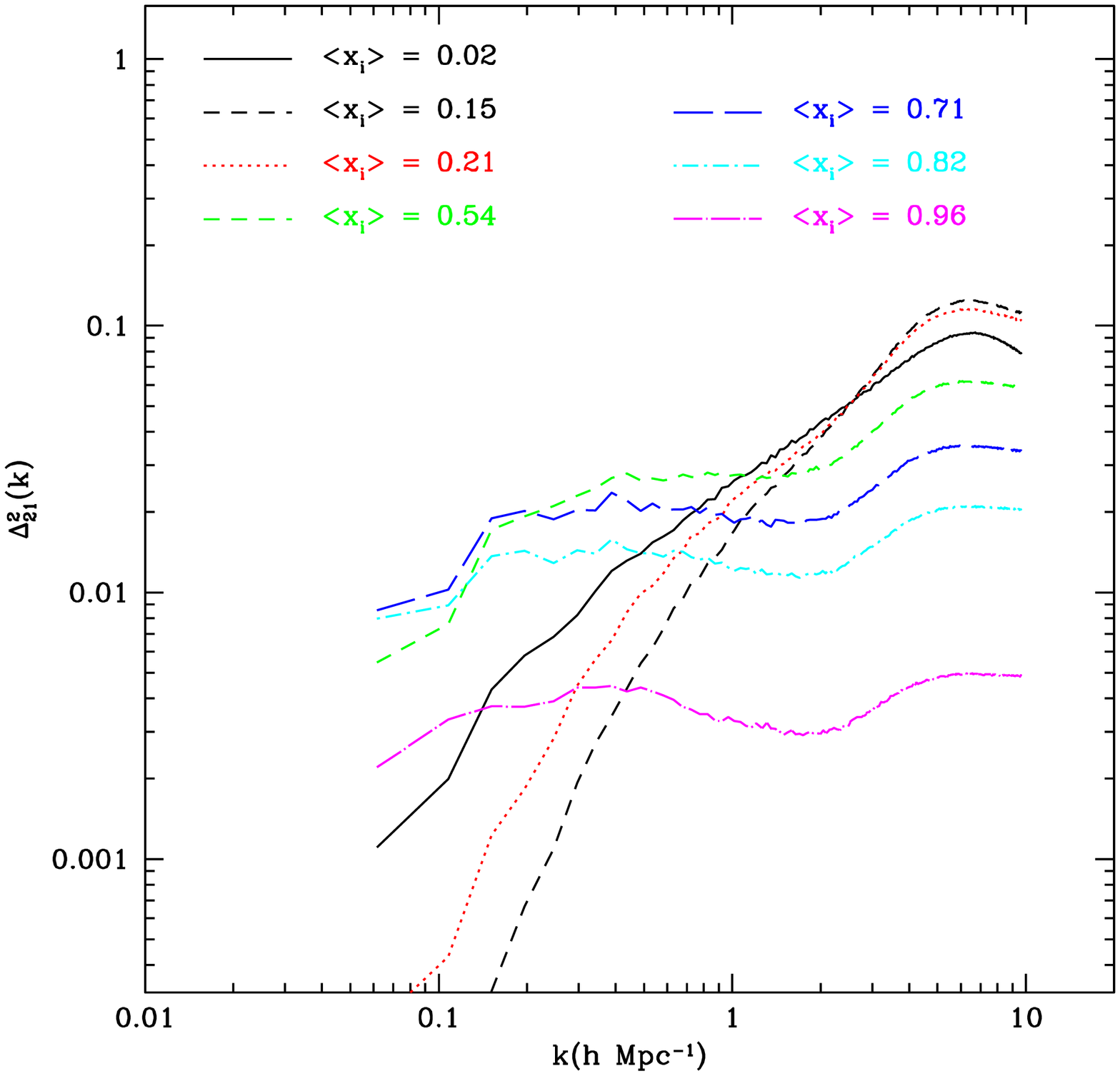}
\includegraphics[scale=0.35]{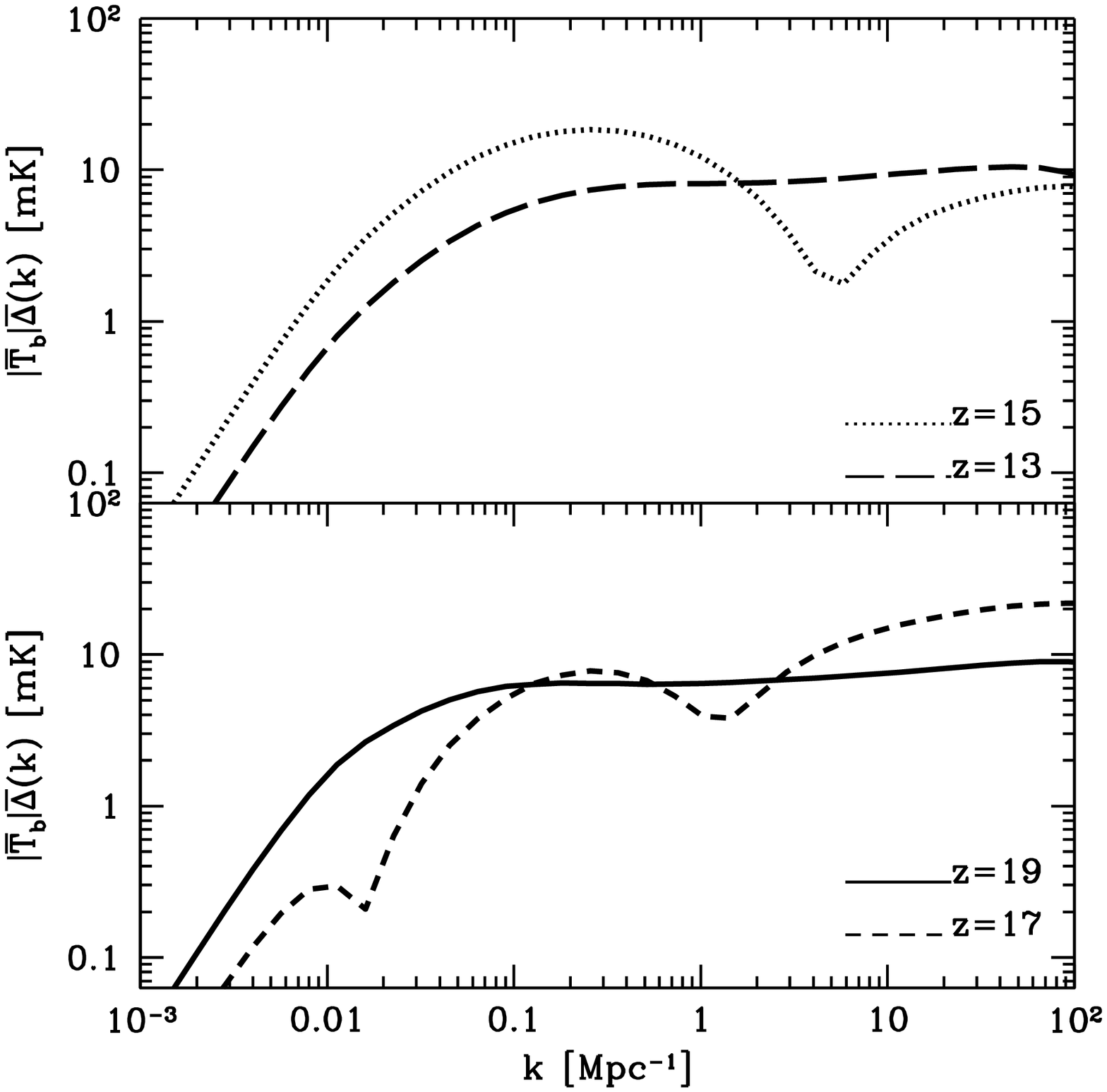}
\caption{\emph{Left panel:}  Evolution of the spherically-averaged 21 cm power spectrum during reionization, from a numerical simulation of that process.  Here the fluctuations are presented in units of $\bdtb^2$ for a fully neutral universe (see eq.~\ref{eq:dtb}); $\VEV{x_i}$ is the ionized fraction during reionization.  From \cite{lidz08-constraint}.  \emph{Right panels:}   Evolution of the spherically-averaged power spectrum \emph{before} reionization begins in earnest, including X-ray heating and Wouthuysen-Field coupling fluctuations.  From \cite{pritchard07}.  In all panels, we use models with parameters similar to the solid curve in Fig.~\ref{fig:global}.  }
\label{fig:pk}
\end{center}
\end{figure}

Observations have provided tantalizing hints about reionization, but even more unanswered questions \citep{fan06-review}.  For example, CMB observations imply that reionization completed by $z \sim 10$ \citep{dunkley08}, but quasar absorption spectra suggest that it may have continued until $z \sim 6$ \cite{fan06}, albeit both with substantial uncertainties.  Both the sky-averaged $\bdtb$ and the 21 cm power spectrum yield much more precise measures of $\bxhi(z)$ (see Fig.~\ref{fig:global} and the left panel of Fig.~\ref{fig:pk}).  The fluctuations briefly fade as galaxies ionize their dense surroundings in the first stage of reionization, then increase as large ionized bubbles form, finally fading again as the gas is ionized.  The principal goal of first-generation experiments (now under construction) is to constrain this time evolution \cite{lidz08-constraint}.

\underline{\it What sources were responsible for reionization?}  The 21 cm sky contains much more information about reionization, because the properties of the ionizing sources strongly affect its topology and have quantifiable impacts on the power spectrum.  Most fundamentally, stars produce well-defined ionized regions, while ``miniquasars" (small accreting black holes) produce much more diffuse features \cite{zaroubi05}. But the 21 cm background can even distinguish between different stellar reionization scenarios \cite{lidz08-constraint}.  For example, more massive galaxies produce larger ionized regions \cite{furl04-bub, mcquinn07}, as shown by the four panels in Fig.~\ref{fig:pics}.  Photoheating (accompanied by a suppression of galaxy formation due to the increased IGM pressure) also slows the growth of bubbles \cite{iliev07-selfreg}.  We can thus indirectly constrain the sources of ionization, including even extremely faint galaxies (which probably dominate the stellar mass budget), and gauge the importance of exotic, very massive Population III stars.  

Of course, exploring this era is a primary goal of many other forthcoming experiments.  Comparison with, e.g., galaxy surveys will provide a clear and complete picture connecting the small-scale physics of galaxies with landmark events encompassing larger scales and illuminate the relation between galaxies and the IGM, if appropriately wide and deep near-IR surveys can be done \cite{lidz09}.  Targeted near-IR follow-up of individual HII regions identified in 21 cm will also identify how reionization proceeds on smaller scales.

\underline{\it How did the cosmic web evolve?}  Once they leave their sources, ionizing photons must propagate through the IGM, where they can be absorbed by dense pockets of neutral gas around galaxies.  These affect the topology of the ionized bubbles during the tail end of reionization, when the pattern of well-defined ionized bubbles disintegrates into the ``cosmic web" that we see in the Ly$\alpha$ forest \cite{furl05-rec, choudhury08}.  This process also imprints distinctive features on the 21 cm signal and allows us to measure directly the evolution of the IGM's structure, as well as the interactions of ionizing sources with the IGM.

\underline{\it Where did the first generations of quasars form, and what were their properties?}  Bright quasars form enormous HII regions in the IGM ($> 40 \Mpc$ across), visible in the 21 cm sky even after the quasar becomes dormant, because the recombination time is relatively long \cite{furl08-fossil}.  Studying these regions in detail will constrain the quasar emission mechanism and their lifetimes, luminosity function, and redshift evolution \cite{wyithe05-qso}.

\underline{\it When did the first galaxies form, and what were their properties?}  As mysterious as they are, the galaxies responsible for reionizing the Universe were probably relatively mature.  The 21 cm transition also probes the very first structures to appear in the Universe -- an era most likely inaccessible even to \emph{JWST}.  The right panel of Fig.~\ref{fig:pk} shows the 21 cm power spectrum during this early era in one typical model.  The first stars produced a ``soft" UV background (below the HI ionization edge) that triggered an absorption feature in the 21 cm background (at $z \approx 20$ in Fig.~\ref{fig:global}).  The background is strongest near these galaxies, so the absorption fluctuations trace their locations, masses, UV spectra, luminosity function and redshift evolution \cite{barkana05-ts}.  The $z=19$ curve in Fig.~\ref{fig:pk} illustrates this era; the fluctuation amplitude can be \emph{stronger} than that during reionization.

\underline{\it When did the first black holes form, and what were their properties?}  Somewhat later, X-rays from the first black holes (and supernovae) heated the Universe (at $z \approx 15$ in Fig.~\ref{fig:global}).  Again, the heating was concentrated around the galaxies hosting these sources, so the observable fluctuations trace their redshift evolution, abundance, masses, and spectra \cite{pritchard07}.  The $z=17$ and 15 curves in Fig.~\ref{fig:pk} show that the sharp contrast between hot and cold regions produces distinctive troughs in the power spectrum, as well as especially large fluctuation amplitudes.  Thus we can securely identify the era in which the first X-ray sources heated the IGM and then follow their importance throughout the reionization era \cite{oh01, ricotti05}.

\underline{\it How does radiative feedback affect high-z galaxy formation?}  Both the UV and X-ray backgrounds are crucial feedback mechanisms in early galaxy formation.  For example, they strongly affect H$_2$, the primary coolant in primordial gas, and therefore regulate both the possible locations of star formation and its end products \cite{haiman00, ahn08}.  Moreover, X-ray heating also affects structure formation and clumping in the IGM \cite{oh03}.  The 21 cm background is the only method to measure  these mechanisms \emph{directly}.  

\section{Milestones} \label{mile}

The ultimate goal of studying the 21 cm background is to make detailed maps of the IGM throughout the ``dark ages" and reionization, as in Fig.~\ref{fig:pics}.  The top axis of Fig.~\ref{fig:global} shows the observed frequency range for these measurements: well within the low-frequency radio regime.  Unfortunately, this is an extremely challenging band, because of terrestrial interference, ionospheric refraction, and (especially) other astrophysical sources (see \cite{furl06-review}).  In particular, the polarized Galactic synchrotron foreground has $T_{\rm sky} \sim 180 (\nu/180 \MHz)^{-2.6} \kel$, at least four orders of magnitude larger than the signal.  For an interferometer, the noise per resolution element (with an angular diameter $\Delta \theta$ and spanning a bandwidth $\Delta \nu$) is then \cite{furl06-review}
\begin{equation}
\Delta T_{\rm noise} \sim 20  \mkel \ \left( \frac{10^4 \sqm}{A_{\rm eff}} \right) \, \left( \frac{10'}{\Delta \theta} \right)^2 \, \left( \frac{1+z}{10} \right)^{4.6} \,  \left( \frac{{\rm MHz}}{\Delta \nu} \, \frac{100 \hr}{t_{\rm int}} \right)^{1/2},
\label{eq:if-sens}
\end{equation}
where $A_{\rm eff}$ is the effective collecting area and $t_{\rm int}$ is the integration time.  These angular and frequency scales correspond to $\sim 30 \Mpc$.  Here we outline the steps required to explore this era in detail, given the challenges implicit in this huge noise.

{\bf The all-sky signal:}  The global background illustrated in Fig.~\ref{fig:global} contains an extraordinary amount of information about the gross properties of galaxies.  These measurements can easily beat down the foreground noise with only a single dipole, so they may provide our first constraints at very high redshifts.  The challenge lies in calibration that is precise enough to extract the signal from instrumental artifacts and the bright foregrounds.  The EDGES experiment \cite{bowman08} has already set upper limits and hopes to measure this signal to $z \sim 20$ over the next decade.

\begin{figure}[htbp]
\begin{center}
\includegraphics[scale=0.35]{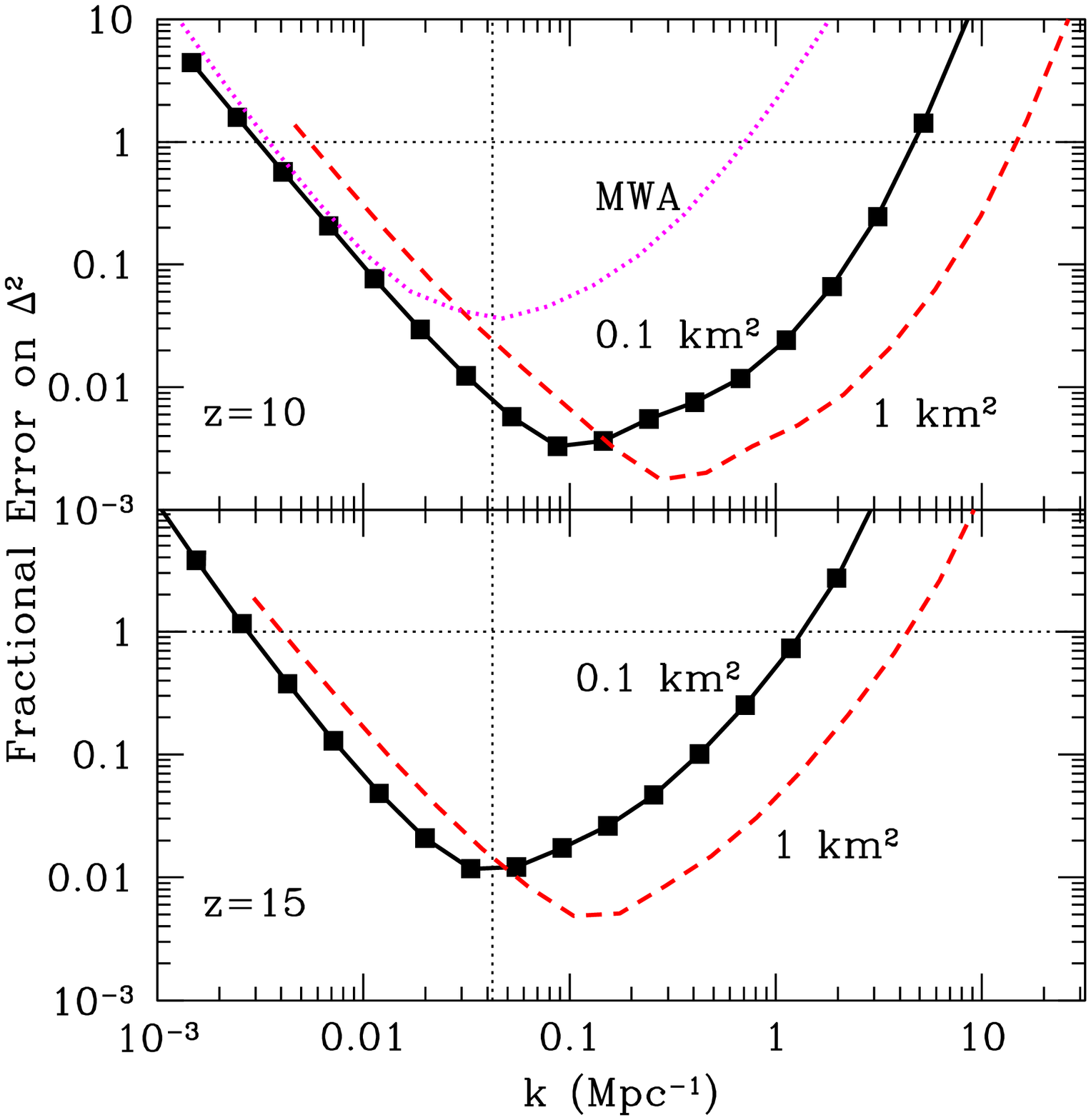}
\includegraphics[scale=0.35]{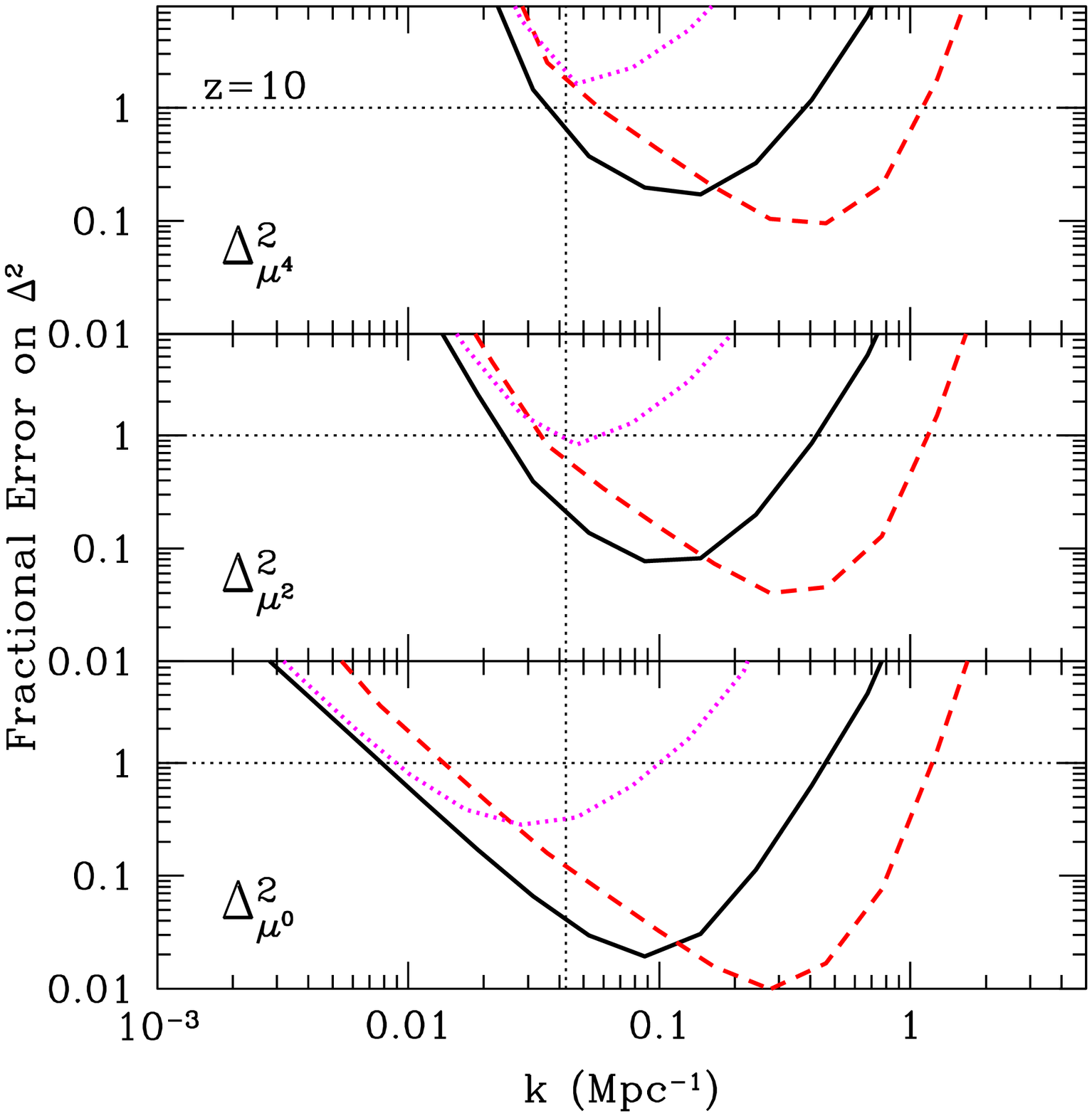}
\caption{\emph{Left panels:}  Sensitivity of some fiducial arrays to the spherically-averaged power spectrum at $z=10$ and $z=15$, (conservatively) assuming a fully neutral emitting IGM.  The solid and dashed curves take $A_{\rm eff}=0.1$ and 1 km$^2$, respectively; in both cases we take  4000 hr of observing split over four fields, a total bandwidth $B=8 \MHz$, $T_{\rm sky} = (471,\,1247) \kel$ at $z=(10,15)$, and $N=5000$ stations centered on a filled core and with an envelope out to $R_{\rm max}=3$~km.  The black squares show the location of the independent $k$ bins in each measurement.  The dotted curve shows the projected MWA sensitivity at $z=10$.  The vertical dotted line corresponds to the bandwidth; modes with smaller wavenumbers are compromised by foreground removal.   \emph{Right panels:} Sensitivity of the same arrays to angular components of the power spectrum at $z=10$.}
\label{fig:sensitivity}
\end{center}
\end{figure}

{\bf First-Generation Arrays:}  The arrays now operating or under construction, including the Murchison Widefield Array (MWA), the Precision Array to Probe the Epoch of Reionization (PAPER), LOFAR, the 21 Centimeter Array, and the Giant Metrewave Radio Telescope, have $A_{\rm eff} \sim 10^4 \sqm$ and so are limited to imaging only the most extreme ionized regions (such as those surrounding bright quasars).  Nevertheless,  these arrays have sufficiently large fields of view ($> 400^{\circ2}$) to make reasonably good statistical measurements \cite{mcquinn06-param, bowman07}.  Fig.~\ref{fig:sensitivity} shows the projected errors for the MWA at $z=10$.  It will be able to detect fluctuations over a limited spatial dynamic range (thanks to foreground removal and thermal noise) and only at $z < 12$, constraining the timing of reionization and some of the source physics.

{\bf Second-Generation Arrays:}  Fig.~\ref{fig:sensitivity} shows that larger telescopes, with $A_{\rm eff} \sim 10^5 \sqm$ (and large fields of view), will clearly be needed for precise measurements, and especially to identify the distinctive features in the power spectrum (such as the flattening at $k > 1 \Mpcinv$ during reionization and the troughs during X-ray heating; see Fig.~\ref{fig:pk}).  Instruments in this class will also be able to measure some more advanced statistics.  This is illustrated in the right panels in Fig.~\ref{fig:sensitivity}, which consider the components of the redshift-space distortions induced by velocity fluctuations.  These are extremely useful for breaking degeneracies in the signal \cite{barkana05-vel} (for more information, see the companion white paper ``Cosmology from the Highly-Redshifted 21 cm Line") but lie beyond the reach of first-generation experiments.

{\bf Imaging Arrays:}  At $A_{\rm eff} \sim 10^6 \sqm$, imaging on moderate scales becomes possible, and statistical constraints become exquisite even at high redshifts  (provided that the large field of view, not strictly necessary for imaging, is maintained; see Fig.~\ref{fig:sensitivity}).  

Plans for these later generations will evolve as we learn more about ``dark age" physics  and the experimental challenges ahead; for example, the Long Wavelength Array and other lower-frequency instruments will study the ionospheric calibration required to explore the high-$z$ regime ($z > 12$, or $\nu < 110 \MHz$) and help determine the relative utility of a terrestrial Square Kilometer Array or a far-side Lunar Radio Array.  At the same time, we must explore whether innovative new telescope designs more closely aligned with the observables, such as the FFT Telescope \cite{tegmark09}, can provide cost-effective improvements.

This roadmap, with accompanying efforts to improve theoretical modeling of the first galaxies and data analysis techniques (such as specialized statistical measures) will ideally position the community to explore the major science questions of high-$z$ galaxy formation.

\newpage

\bibliographystyle{apj}
\bibliography{Ref_composite}

\end{document}